\documentclass[10pt,
 amsmath,amssymb,
floatfix,
lengthcheck,%
]{revtex4-1}
\usepackage{graphicx}
\usepackage{graphics}
\usepackage{epstopdf} 
\usepackage{dcolumn}
\usepackage{bm}

\begin{document}

\title{Effect of solute segregation on shear-induced grain boundary motion}
\author{Changjian Wang}
\affiliation{Group for Simulation and Theory of Atomic-Scale Material Phenomena (stAMP), Department of Mechanical and Industry Engineering, Northeastern University, Boston, Massachusetts 02115, USA.}
\author{Moneesh Upmanyu}
\email{mupmanyu@neu.edu}
\affiliation{Group for Simulation and Theory of Atomic-Scale Material Phenomena (stAMP), Department of Mechanical and Industry Engineering, Northeastern University, Boston, Massachusetts 02115, USA.}
\affiliation{Department of Bioengineering, Northeastern University, Boston, Massachusetts 02115, USA.}

\begin{abstract}
Atomic-scale simulations are performed to study the effect of solute segregation on the shear-induced motion of select grain boundaries in the classical $\alpha$-Fe/C system. 
At shear rates larger than the solute diffusion rate, we observe a transition from coupled motion to sliding.
Below a critical solute excess,
the boundaries break away from the solute cloud and move in a coupled motion.  At smaller shear rates,
we observe extrinsic coupled motion at small stresses indicating that the coupling is aided by convective solute diffusion along the boundary. Our studies underscore the role of solutes in modifying the bicrystallography, temperature and rate dependence of shear accommodation at grain boundaries. 
\end{abstract}

\maketitle
The mechanics of polycrystals is shaped by the underlying grain boundaries. It is well established that the microstructure evolves in response to external stresses via motion of the constituent boundaries~\cite{book:SuttonBalluffi:1995, *book:GottsteinShvindlerman:1999}. As a consequence, tailoring the polycrystal properties to the desired application - be it high strength, ductility, or toughness - is no longer limited to the processing routes; it requires a fundamental understanding of the motion of grain boundaries to the locally generated stress state. Of note is the response to shear as the relatively weaker grain boundary structure renders them susceptible to sliding, i.e. $v_\parallel=S\tau$, where $S$ is the sliding constant associated with the tangential velocity $v_\parallel$ in response to a shear stress $\tau$. There is growing evidence that the shear can also couple to the normal motion $v_n$~\cite{gb:CahnTaylor:2004, gb:GutkinOvidko:2005}. The coupling constant, defined as the ratio $v_{||}/v_n=\beta$ for fully coupled motion ($S=0$), is sensitive to the boundary type, temperature and shear rate, as recently validated by experiments and atomic-scale computations~\cite{gb:CahnMishin:2006, *gbm:ZhangDuSrolovitz:2008, *gbm:IvanovMishin:2008, *gbm:GorkayaGottstein:2010, *gbm:MompiousCaillard:2011, *gb:KarmaTrauttMishin:2012}.

Current understanding of this interplay between shear stress and boundary motion is based on studies on elemental systems; the effect of solutes (or impurities) and related defects has been largely ignored. These extrinsic effects are important in both commercial purity metals and engineered alloy systems as the solutes preferentially segregate to the grain boundaries~\cite{gbseg:Seidman:2002}. The modified boundary energetics and kinetics can lead to deviations from the theorized trends. As an example, the effect on normal motion has been studied extensively, dating back to early work by Cahn, L\"ucke and St\"uwe~\cite{imdrag:LuckeDetert:1957, *imdrag:Cahn:1962, *gbm:LuckeStuwe:1971}, wherein the solutes exert a drag when the boundary moves with the solute cloud - the loaded regime. 

The effect of solute segregation on shear accommodation at the grain boundaries remains unexplored. We address this vacuum by performing atomic-scale simulations on flat grain boundaries in BCC-iron, decorated with an equilibrated interstitial carbon segregation profile and subject to an imposed shear. We explore the effect of (boundary) solute excess $\Gamma_i$, shear-rate, $\dot{\gamma}$ and temperature $T$ on the boundary deformation, viz. sliding versus coupled motion. The response is self-consistently compared with that of a pure boundary. 

\begin{figure}[htp]
\includegraphics[width=\columnwidth]{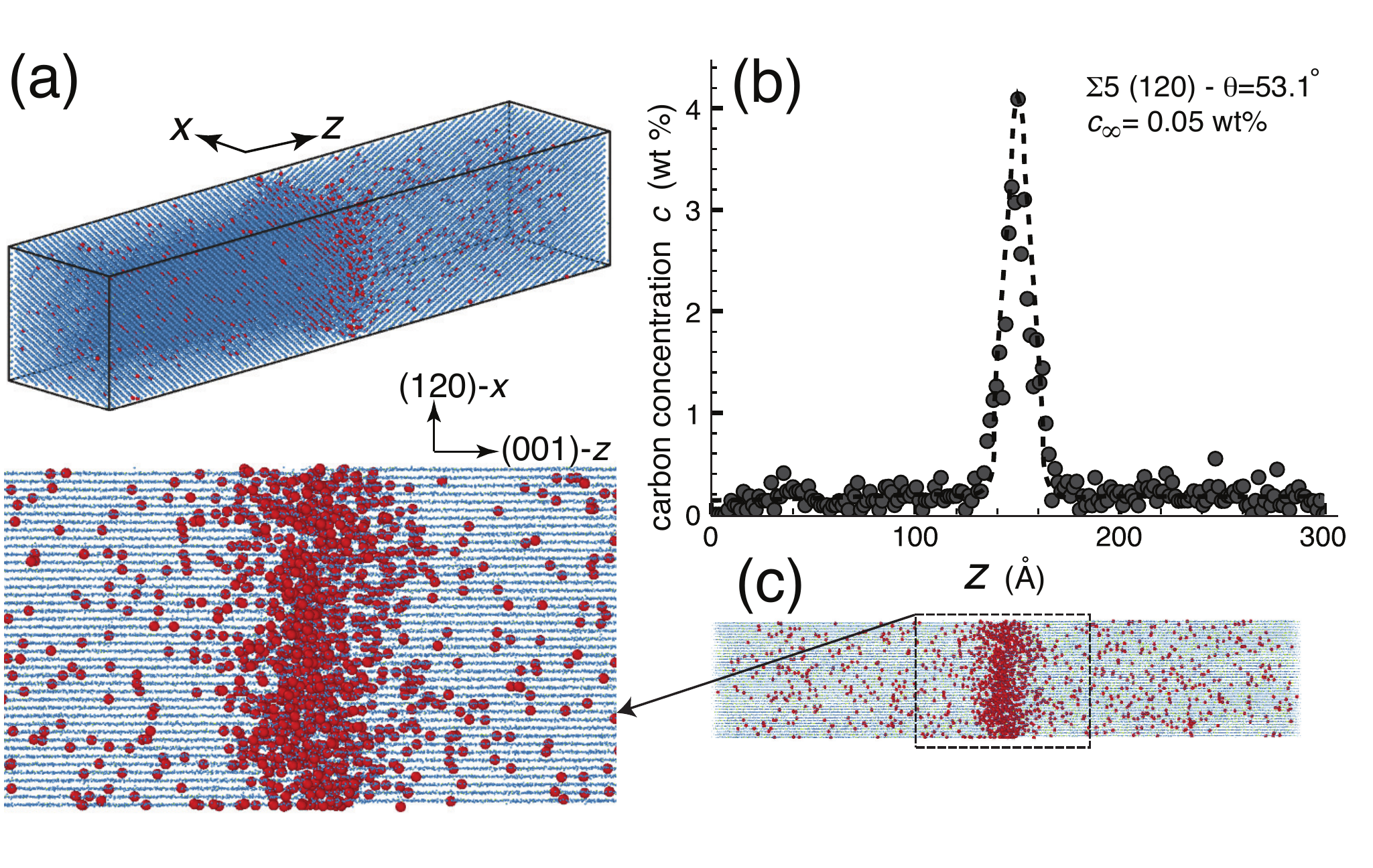}
 \caption{(color online) (a) Atomic bicrystal configuration of an equilibrated $\Sigma5\,(001)$-$\theta=53.1^\circ$ grain boundary in BCC-Fe. The bulk impurity concentration is $c_\infty=0.05$\,wt\%. Here and elsewhere, blue and red atoms denote Fe and C, respectively. The Fe atom size is reduced to better depict the distribution of C atoms and their segregation at the boundary. (b) The segregation profile $c(z)$ at $T=550$\,K that follows the dilute approximation $c\approx c_\infty\exp(-U/k_BT)$, where $U\approx U(z)$ is the solute-boundary interaction energy . 
\label{fig:figure1}
 }
\end{figure}

Figure~\ref{fig:figure1}a shows the atomic configuration of one of the symmetric tilt grain boundaries in $\alpha$-Fe, a  high-angle $\Sigma5\,(001)$-$\theta=53.1^\circ$. The bulk carbon concentration is fixed at $c_\infty=0.05$\,wt\% and the segregation at the grain boundary is clearly visible. Empirical interatomic potentials of the Hepburn-Ackland variety are used for describing the Fe-Fe and Fe-C interactions~\cite{intpot:HepburnAckland:2008}. The simulation cell used to extract the pure boundary structure is periodic in-plane ($L_x=6.1$\,nm and $L_y=6.2$\,nm) and terminates at free surfaces normal to the boundary ($L_z=30.4$\,nm). The structure is locally relaxed at $0\,$K  using conjugate gradient minimization and the temperature is gradually increased at zero pressure via isothermal-isobaric (NPT) molecular dynamic (MD) simulations (Nose-Hoover thermostat, time step of 2\,fs). Volume perturbations normal to the boundary plane are employed to finally equilibrate the structure for $0.5$\,ns at the desired temperature.

Equilibrium carbon segregation profile is obtained using grand-canonical Monte-Carlo (GCMC) computations~\cite{atsim:FrenkelSmit:2002}. To eliminate spurious effects due to segregation at the free surfaces, we employ fixed $z$-edge regions which are eventually  relaxed to $c_\infty$.
We limit the study to low carbon concentrations ($<0.2$\,wt\%), and for efficiency the Fe-atoms are frozen and interstitial carbon atoms are added, deleted and moved in accordance with the prescribed chemical potential. The addition of carbon atoms leads to changes in pressure, and to that end NPT MD is used to adjust the cell volume, periodically and towards the end of the simulation. The equilibrated profile, averaged over $x$-$y$ planes and plotted in Fig.~\ref{fig:figure1}b, shows maximum at the boundary with 
an impurity excess $\Gamma_i\approx27$\,atoms/nm$^2$ distributed over a  boundary width $\delta\approx4$\,nm (Fig.~\ref{fig:figure1}c). Canonical (NVT) MD simulations are employed to study the shear response. One of the two $z$-edges is held fixed while the atoms in the other $z$-edge are prescribed a constant shear rate $\dot{\gamma}$. A local orientation order parameter~\cite{gbm:TrauttUpmanyuKarma:2006} together with the location of the highest peak in the segregation profile $c(z)$ is used to track the boundary dynamically. 
\begin{figure}[bth]
\includegraphics[width=0.8\columnwidth]{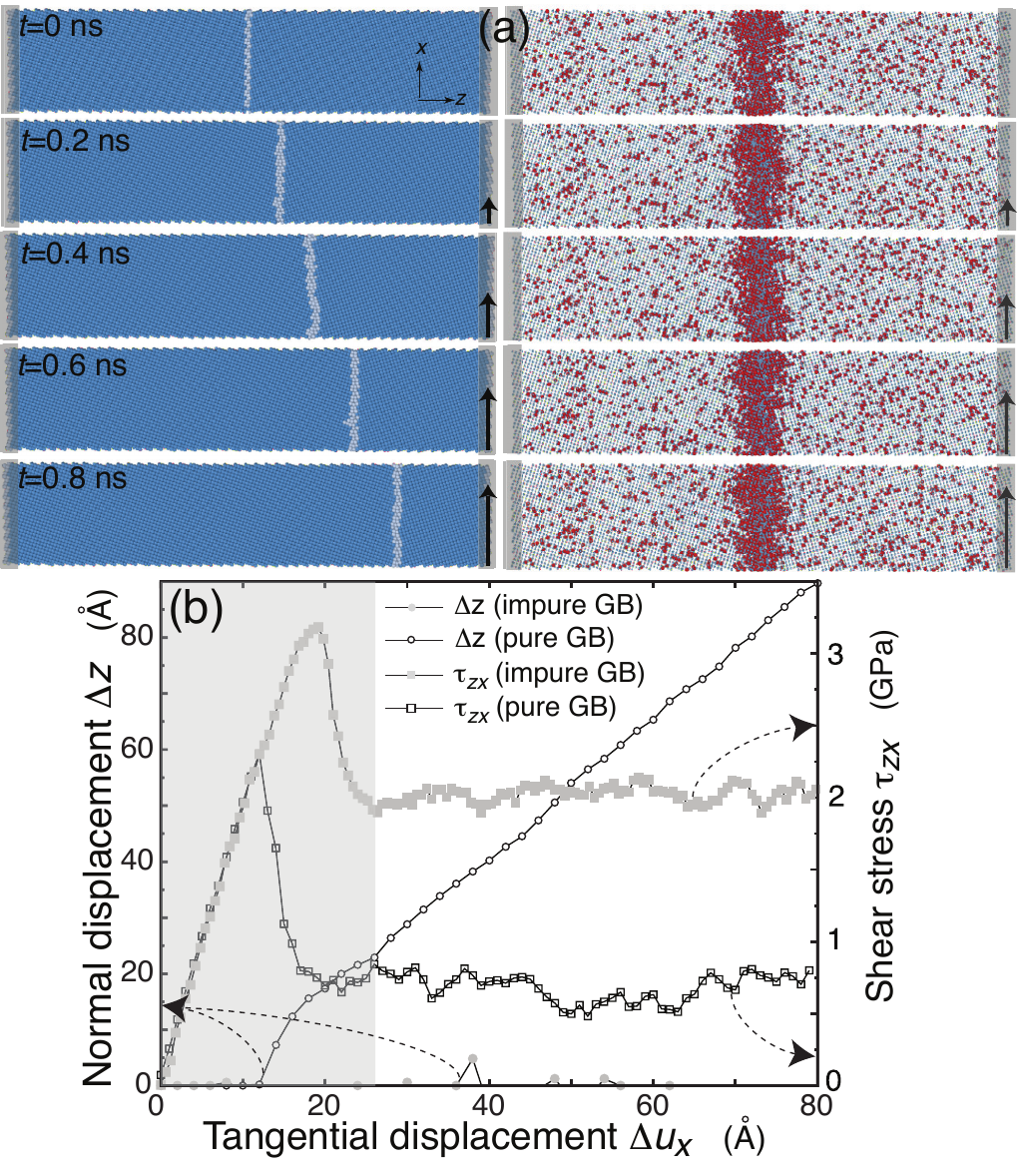}
 \caption{(color online) (a) Atomic configurations of the (left) pure and (right) loaded bicrystals in Fig.~\ref{fig:figure1} at shear rate $\dot{\gamma}=1$\,m/s. The vertical arrows at the $z$-edges indicate the shear displacement. The boundary atoms are shaded light gray. (b) Evolution of the normal boundary displacement $\Delta z$ and the shear stress $\tau_{zx}$ with the tangential displacement $\Delta u_x$ for pure (open symbols) and loaded  (filled symbols) boundary. \label{fig:figure2}}
\end{figure}

Figure~\ref{fig:figure2} shows the atomic configurations of the pure and segregated boundary at temperature $T=550$\,K and shear rate $\dot{\gamma}=1$\,m/s (strain rate $\approx10^{7}$\,s$^{-1}$) that is higher than the diffusion rate, $D(550\,{\rm K})/\langle r \rangle\approx10^{-2}-10^{-4}$\,m/s. Here, $\langle r \rangle$ is the interstitial diffusion length of the order of the Fe lattice parameter $a$. Following an initial transient, the pure boundary exhibits fully coupled motion with a coupling constant $\beta\approx1$ that is in excellent agreement with geometrical predictions~\cite{gb:CahnTaylor:2004}, i.e. $\beta=2\tan\theta/2$. It then follows that the normal driving force $p$ due to the orientation dependent elastic anisotropy is negligible\cite{gbm:Schonfelder:1997, *gbm:ZhangMendelev:2004}.  The shear stress  $\tau_{zx}\approx0.75$\,GPa and devoid of serrations associated with stick-slip behavior ($S\approx0$),  consistent with past studies on
fully coupled boundaries~\cite{gb:CahnMishin:2006}. 
\begin{figure}[htp]
\includegraphics[width=\columnwidth]{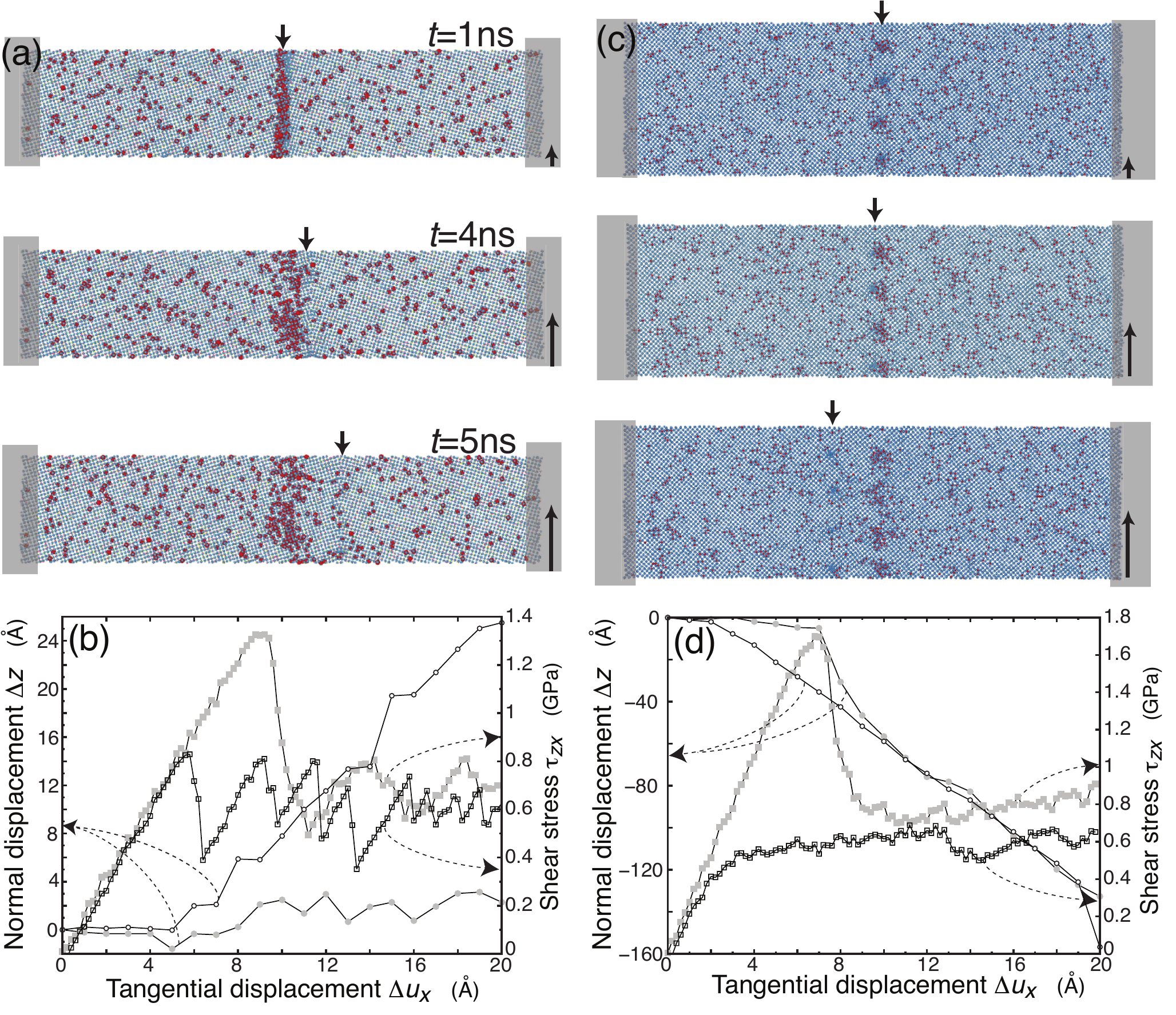}
 \caption{(a) Atomic configurations of the $\Sigma5$ bicrystal with $c_\infty=0.1$\,wt\% and subject to shear rate $\dot{\gamma}=1$\,m/s at $T=1000$\,K. The location of the boundary is indicated by the vertical arrow. (b) Eevolution of normal displacement and shear stress, as in Fig.~\ref{fig:figure2}b. (c-d) Same as in (a) and (b), but for a $\theta=7.2^\circ$ low-angle grain boundary with $c_\infty=0.18$\,wt\%. \label{fig:figure3}}
\end{figure}

The response of the solute-loaded boundary is dramatically different. As in the pure case, the boundary exhibits sliding during the initial transient as the stress increases and then plateaus to a steady-state value, $\tau_{zx}\approx2$\,GPa. Thereafter, the boundary continues to slide with negligible normal motion - the motion is uncoupled ($\beta\approx0$). The shear stress is higher due to the segregation yet the transition to sliding suggests that the solute drag to normal motion exerted by the mostly static solute cloud~\cite{imdrag:LuckeDetert:1957, *imdrag:Cahn:1962, *gbm:LuckeStuwe:1971}, $p_i^z\approx N \int_{\partial\mathcal{S}} c\nabla_z U\,dz$,
effectively suppresses the coupled motion. Here, $N$ is the number density of solute sites within the grain boundary region $\partial\mathcal{S}$ and the spatial gradient $\nabla_z U$ is the drag force per solute. Evidently, the boundary is weaker with respect to sliding due to a combination of segregation energetics and associated structural transitions that can lubricate the boundary, as well as solute diffusion through the boundary that can be itself aided by the stress~\cite{gb:CahnLarche:1978}. 

Decreasing the solute concentration and increasing the temperature results in another class of behavior wherein the boundary breaks away from the solute cloud and transitions to the unloaded, shear-coupled regime. Two instances of the transition at $T=1000$\,K are shown in Fig.~\ref{fig:figure3}: the $\Sigma5$ boundary at $c_\infty=0.02$\,wt\% and a low-angle $\theta=7.2^\circ(001)$ boundary at $c_\infty=0.05$\,wt\%. The shear rate is same as in Fig.~\ref{fig:figure2}, and although the carbon diffusivity is enhanced by orders of magnitude [$D(1000\,{\rm K})\approx10^{-10}$m$^2$/s], it is still high compared to the diffusion rate $D/\langle r \rangle\approx0.1$\,m/s. The reduced carbon concentration translates to much smaller impurity excess, and in turn a smaller (normal) impurity drag. As the bicrystal increases its shear stress in response to the shear displacement (Fig.~\ref{fig:figure3}b), the coupled response is constrained. The stress induces a bulk driving pressure normal  to the boundary that scales with the elastic anisotropy between the two crystals~\cite{gb:CahnTaylor:2004, gbm:Schonfelder:1997, *gbm:ZhangMendelev:2004}. Since the elastic anisotropy in Fe is weak, the boundary slides to absorb the imposed shear at low shear displacement and stresses. At large stresses, $p>p_i^z$ and the boundary  breaks away from the solute cloud. The shear stress decreases sharply and then plateaus as the boundary transitions to the unloaded, coupled regime ($p\approx0$).  

\begin{figure*}[htp]
\includegraphics[width=1.75\columnwidth]{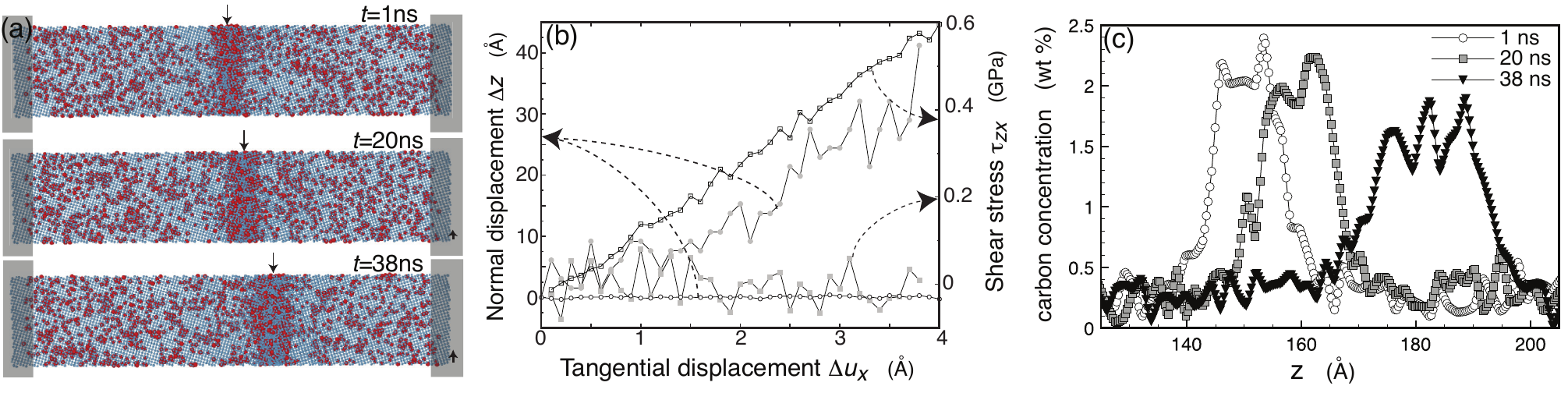}
 \caption{(a) Atomic plots showing the coupled motion of the segregated $\Sigma5$ boundary at $T=1000$\,K and $\dot{\gamma}=0.01$\,m/s. The bulk carbon concentration is $c_\infty=0.07$\,wt\%. (b) The normal displacement and shear stress evolution plot as in Fig.~\ref{fig:figure2}b, and (c) the corresponding boundary segregation profiles. \label{fig:figure4}}
\end{figure*}
The shear stress for the unloaded boundary is higher 
as its motion is modified by the small fraction of solute atoms that continually accumulate at the boundary as it moves through the solid solution. In the case of the $\Sigma5$ boundary, the normal displacement of the pure boundary exhibits a staircase-like profile with a correspondingly serrated shear stress profile, characteristic of a mixed mode that involves both coupling and sliding under these loading conditions. The unloaded boundary is weakly coupled suggesting that the drag force due to relatively static isolated impurities is significant~\cite{imdrag:MendelevSrolovitz:2002, *imdrag:MendelevSrolovitzAckland:2005, *imdrag:ZepedaRuizGilmer:2006}. Interestingly, the normal motion consists of small intervals characterized by negative normal motion that indicate that the motion is jerky. The vacated solute cloud itself acts as an additional bulk driving force that can further modify the normal motion. The shear stress evolution is also non-uniform.  Eventually, the boundary accumulates enough solutes and transitions back to the loaded regime. Over experimental time-scales, we expect the overall motion to be quite non-uniform with cyclic shear stress due to transitions between the loaded and unloaded motion. Our results show that for fixed loading conditions the transition between sliding and breakaway is associated with a critical solute excess $\Gamma^\ast$.

The response of the low-angle boundary shown in Fig.~\ref{fig:figure3}c-d is similar. The Cottrell atmosphere of solutes around the boundary dislocations is clearly visible. The pure boundary is fully coupled at this temperature with negligible sliding, as expected for a long-angle boundary. Beyond a critical tangential displacement, the dislocations escape from the respective Cottrell atmospheres. The subsequent normal motion takes place via dislocation glide on parallel slip planes along the boundary normal. The shear stress again decreases sharply and plateaus to a value slightly larger than the pure boundary as the dislocations move through the solid solution. Unlike the $\Sigma5$ boundary, the coupled motion is quantitatively similar to the pure boundary due to the reduced frequency of interactions between the dislocation cores and the solutes. The low-angle boundary does not exhibit any transition to sliding at higher shear rates or temperatures as it is strongly coupled to almost the bulk melting point. 

Lowering the shear rate results in a transition to shear coupled behavior. Figure~\ref{fig:figure4}a shows the response of the $\Sigma5$ boundary for $\dot{\gamma}=0.01$\,m/s, $c_\infty=0.07$\,wt\% and $T=1000$\,K. The atomic configurations (Fig.~\ref{fig:figure4}a) reveal that the solute cloud is now quite mobile relative to the boundary motion as the bulk carbon diffusion rate is almost an order of magnitude faster. We immediately see a transition to shear-coupled behavior. The normal displacement plotted in Fig.~\ref{fig:figure4}b is quite sensitive to the applied shear - the boundary displaces by almost $\Delta z\approx4$\,nm for a tangential displacement of $\Delta u_x\approx0.4$\,nm yielding an effective coupling constant of $v_\parallel/v_n\approx0.1$. Under identical conditions and for the same range of tangential displacement, the pure boundary shows negligible normal motion, indicating that the coupling constant is smaller. Comparison with the mixed response that occurs over larger tangential displacements (Fig.~\ref{fig:figure3}b) confirms that the coupling is almost an order of magnitude smaller, underscoring the effect of solute-boundary interactions. 

The nature of the boundary motion and the corresponding shear stress evolution is also quite different (Fig.~\ref{fig:figure4}b). As before, the boundary motion is jerky and the coupling is non-uniform with intervals where it is negative. A key aspect of the motion is that unlike the pure or the unloaded grain boundary, the coupled motion occurs at a much smaller stress that fluctuates and changes sign. However, the boundary does not wander~\cite{gbm:TrauttUpmanyuKarma:2006}, as its motion remains correlated to the shear stress. Evidently, the solute cloud aids boundary motion by lowering the shear stress that drives the coupling behavior. 

We delegate a detailed mechanistic treatment to a subsequent study and for now consider the case where at steady-state the solute cloud moves normal and tangential to the boundary plane to accommodate the shear. Combining the simplified CLS model for solute drag with the Cahn-Taylor model for coupled motion~\cite{gb:CahnTaylor:2004}, 
\begin{align}
&v_{\parallel}=S\tau + \frac{\beta}{\alpha c_\infty} \left(p + \beta\tau\right),\quad
v_{n} = \frac{1}{\alpha c_\infty} \left( p + \beta\tau\right),\\
&\text{where}\,\,\alpha=4N k_BT \int_{-\delta/2}^{\delta/2} \frac{dz}{D_\parallel} \sinh^2 \left[ \frac{U(z)}{k_BT} \right]\nonumber
\end{align}
is a parameter related to the solute-impurity interaction. We assume that the intrinsic normal mobility $M$ in the low velocity limit is negligible compared to the extrinsic mobility$M_i\approx1/\alpha c_\infty$~\cite{gbm:Kappes:2008}. In the $p=0$ limit, we get $v_\parallel/v_n=\beta + S\alpha c_\infty/\beta$ which still cannot reconcile the large decrease in $v_\parallel/v_n$ that we observe. 

While we cannot rule out additional sources of normal driving forces on the segregated boundary\footnote{Elastic anisotropy can be important for the loaded boundary as the chemical potential of interstitial carbon is sensitive to stress distribution within the grain boundary region}, a key limitation is that the CLS-based solute drag models ignore the convective solute flux along the boundary~\cite{imdrag:RoyBauer:1975, *imdrag:MendelevSrolovitzWE:2001, *imdrag:KorzhenevskiiBauschSchmitz:2006}. The evolution of the concentration profile $c(z)$ plotted in Fig.~\ref{fig:figure4}c shows that the solute diffusion is significant; the boundary widens with an additional peak  in the segregation profile, and since the boundary diffusivity of interstitial carbon is much higher than in the bulk $D_\parallel\gg D$, this is suggestive of boundary motion-induced solute diffusion that is analogous to the phenomenon of diffusion-induced grain boundary motion (DIGM)~\cite{digm:CahnPanBalluffi:1979, *digm:HillertPurdy:1978, *digm:Handwerker:1988, *digm:Penrose:2004}. The diffusion gradient is associated with a tangential force $\partial_xp^x_i=k_BT \partial_x(c-c_\infty)$. As an example, assuming the boundary is liquid-like with a viscosity $\eta$ and the boundary diffusion is uncorrelated with the normal motion [$c(z)\approx c^{eq}(z)$ and $\partial_zp_i^z\approx c \nabla_z U(z)$], a simple hydrodynamic balance within the boundary region relates the solute diffusion to an extrinsic slip $v_\parallel^\prime$,
\begin{align}
k_BT \partial_x(c-c_\infty) = \partial_y(\eta\partial_y v_\parallel^\prime). 
\end{align}
The generated slip aids the imposed stress and therefore lowers its value, consistent with our simulations.  

In summary, we find that shear stress driven boundary motion is sensitive to the shear rate relative to the solute diffusion time scale $D(T)/\langle r \rangle$. At large shear rates $\dot{\gamma}\gg D/\langle r \rangle$, boundaries that are otherwise fully coupled undergo a transition to pure sliding, or break away from the solute cloud to preserve the coupled response. The nature of the transition is controlled by the solute excess: at low concentrations, the boundaries favor breakaway to the unloaded regime, and high concentrations result in sliding. For $\dot{\gamma}\sim D/\langle r \rangle$, i.e. low shear rate or high temperatures, the solute cloud can move with the boundary. The coupled behavior occurs at significantly lower stress as it is modified by normal and tangential solute flux, and the effective coupling constant can different markedly from that predicted by geometric considerations. In the case of high-angle boundaries, we observe a decrease in the coupling by almost an order of magnitude. The observations underscore the importance of solute segregation in determining both the nature and extent of the shear deformation of grain boundaries.

This work has been supported in part by the Office of Naval Research Structural Metallics Program (N000141010866), and the US Army Armament Research, Development and Engineering Center (ARDEC). 

%


\begin{thebibliography}{10}%
\makeatletter
\providecommand \@ifxundefined [1]{%
 \ifx #1\undefined \expandafter \@firstoftwo
 \else \expandafter \@secondoftwo
\fi
}%
\providecommand \@ifnum [1]{%
 \ifnum #1\expandafter \@firstoftwo
 \else \expandafter \@secondoftwo
\fi
}%
\providecommand \enquote [1]{``#1''}%
\providecommand \bibnamefont  [1]{#1}%
\providecommand \bibfnamefont [1]{#1}%
\providecommand \citenamefont [1]{#1}%
\providecommand\href[0]{\@sanitize\@href}%
\providecommand\@href[1]{\endgroup\@@startlink{#1}\endgroup\@@href}%
\providecommand\@@href[1]{#1\@@endlink}%
\providecommand \@sanitize [0]{\begingroup\catcode`\&12\catcode`\#12\relax}%
\@ifxundefined \pdfoutput {\@firstoftwo}{%
 \@ifnum{\z@=\pdfoutput}{\@firstoftwo}{\@secondoftwo}%
}{%
 \providecommand\@@startlink[1]{\leavevmode}%
 \providecommand\@@endlink[0]{}%
}{%
 \providecommand\@@startlink[1]{%
  \leavevmode
  \pdfstartlink
   attr{/Border[0 0 1 ]/H/I/C[0 1 1]}%
   user{/Subtype/Link/A<</Type/Action/S/URI/URI(#1)>>}%
  \relax
 }%
 \providecommand\@@endlink[0]{\pdfendlink}%
}%
\providecommand \url  [0]{\begingroup\@sanitize \@url }%
\providecommand \@url [1]{\endgroup\@href {#1}{\urlprefix}}%
\providecommand \urlprefix [0]{URL }%
\providecommand \Eprint[0]{\href }%
\@ifxundefined \urlstyle {%
  \providecommand \doi [1]{doi:\discretionary{}{}{}#1}%
}{%
  \providecommand \doi [0]{doi:\discretionary{}{}{}\begingroup
  \urlstyle{rm}\Url }%
}%
\providecommand \doibase [0]{http://dx.doi.org/}%
\providecommand \Doi[1]{\href{\doibase#1}}%
\providecommand \selectlanguage [0]{\@gobble}%
\providecommand \bibinfo [0]{\@secondoftwo}%
\providecommand \bibfield [0]{\@secondoftwo}%
\providecommand \translation [1]{[#1]}%
\providecommand \BibitemOpen[0]{}%
\providecommand \bibitemStop [0]{}%
\providecommand \bibitemNoStop [0]{.\EOS\space}%
\providecommand \EOS [0]{\spacefactor3000\relax}%
\providecommand \BibitemShut [1]{\csname bibitem#1\endcsname}%
\bibitem{book:SuttonBalluffi:1995}%
  \BibitemOpen
  \bibfield{author}{%
  \bibinfo {author} {\bibfnamefont{A.~P.}\ \bibnamefont{Sutton}}\ and\ \bibinfo
  {author} {\bibfnamefont{R.~W.}\ \bibnamefont{Balluffi}},\ }%
  \emph{\bibinfo {title} {Interfaces in crystalline materials}}\ (\bibinfo
  {publisher} {Clarendon Press},\ \bibinfo {year} {1995})\BibitemShut{NoStop}%
\bibitem{book:GottsteinShvindlerman:1999}%
  \BibitemOpen
  \bibfield{author}{%
  \bibinfo {author} {\bibfnamefont{G.}~\bibnamefont{Gottstein}}\ and\ \bibinfo
  {author} {\bibfnamefont{L.~S.}\ \bibnamefont{Shvindlerman}},\ }%
  \emph{\bibinfo {title} {Grain boundary migration in metals: {T}hermodynamics,
  kinetics, applications}}\ (\bibinfo {publisher} {CRC Press},\ \bibinfo {year}
  {1999})\BibitemShut{NoStop}%
\bibitem{gb:CahnTaylor:2004}%
  \BibitemOpen
  \bibfield{author}{%
  \bibinfo {author} {\bibfnamefont{J.~W.}\ \bibnamefont{Cahn}}\ and\ \bibinfo
  {author} {\bibfnamefont{J.~E.}\ \bibnamefont{Taylor}},\ }%
  \bibfield{journal}{%
  \bibinfo {journal} {Acta Mater.}\ }%
  \textbf{\bibinfo {volume} {52}},\ \bibinfo {pages} {4887} (\bibinfo {year}
  {2004})\BibitemShut{NoStop}%
\bibitem{gb:GutkinOvidko:2005}%
  \BibitemOpen
  \bibfield{author}{%
  \bibinfo {author} {\bibfnamefont{M.~Y.}\ \bibnamefont{Gutkin}}\ and\ \bibinfo
  {author} {\bibfnamefont{I.~A.}\ \bibnamefont{{Ovid'ko}}},\ }%
  \bibfield{journal}{%
  \Doi{10.1063/1.2147721}{\bibinfo {journal} {Appl. Phys. Lett.}}\ }%
  \textbf{\bibinfo {volume} {87}},\ \bibinfo {eid} {251916} (\bibinfo {year}
  {2005})\BibitemShut{NoStop}%
\bibitem{gb:CahnMishin:2006}%
  \BibitemOpen
  \bibfield{author}{%
  \bibinfo {author} {\bibfnamefont{J.~W.}\ \bibnamefont{Cahn}}, \bibinfo
  {author} {\bibfnamefont{Y.}~\bibnamefont{Mishin}},\ and\ \bibinfo {author}
  {\bibfnamefont{A.}~\bibnamefont{Suzuki}},\ }%
  \bibfield{journal}{%
  \bibinfo {journal} {Acta Mater.}\ }%
  \textbf{\bibinfo {volume} {54}},\ \bibinfo {pages} {4953} (\bibinfo {year}
  {2006})\BibitemShut{NoStop}%
\bibitem{gbm:ZhangDuSrolovitz:2008}%
  \BibitemOpen
  \bibfield{author}{%
  \bibinfo {author} {\bibfnamefont{H.}~\bibnamefont{Zhang}}, \bibinfo {author}
  {\bibfnamefont{D.}~\bibnamefont{Du}},\ and\ \bibinfo {author}
  {\bibfnamefont{D.~J.}\ \bibnamefont{Srolovitz}},\ }%
  \bibfield{journal}{%
  \bibinfo {journal} {Phil. Mag.}\ }%
  \textbf{\bibinfo {volume} {88}},\ \bibinfo {pages} {243} (\bibinfo {year}
  {2008})\BibitemShut{NoStop}%
\bibitem{gbm:IvanovMishin:2008}%
  \BibitemOpen
  \bibfield{author}{%
  \bibinfo {author} {\bibfnamefont{V.~A.}\ \bibnamefont{Ivanov}}\ and\ \bibinfo
  {author} {\bibfnamefont{Y.}~\bibnamefont{Mishin}},\ }%
  \bibfield{journal}{%
  \bibinfo {journal} {Phys. Rev. B}\ }%
  \textbf{\bibinfo {volume} {78}},\ \bibinfo {pages} {064106} (\bibinfo {year}
  {2008})\BibitemShut{NoStop}%
\bibitem{gbm:GorkayaGottstein:2010}%
  \BibitemOpen
  \bibfield{author}{%
  \bibinfo {author} {\bibfnamefont{T.}~\bibnamefont{Gorkaya}}, \bibinfo
  {author} {\bibfnamefont{T.}~\bibnamefont{Burlet}}, \bibinfo {author}
  {\bibfnamefont{D.~A.}\ \bibnamefont{Molodov}},\ and\ \bibinfo {author}
  {\bibfnamefont{G.}~\bibnamefont{Gottstein}},\ }%
  \bibfield{journal}{%
  \bibinfo {journal} {Scripta Mat.}\ }%
  \textbf{\bibinfo {volume} {63}},\ \bibinfo {pages} {633} (\bibinfo {year}
  {2010})\BibitemShut{NoStop}%
\bibitem{gbm:MompiousCaillard:2011}%
  \BibitemOpen
  \bibfield{author}{%
  \bibinfo {author} {\bibfnamefont{F.}~\bibnamefont{Mompiou}}, \bibinfo
  {author} {\bibfnamefont{M.}~\bibnamefont{Legros}},\ and\ \bibinfo {author}
  {\bibfnamefont{D.}~\bibnamefont{Caillard}},\ }%
  \bibfield{journal}{%
  \bibinfo {journal} {J. Mater. Res.}\ }%
  \textbf{\bibinfo {volume} {46}},\ \bibinfo {pages} {4308} (\bibinfo {year}
  {2011})\BibitemShut{NoStop}%
\bibitem{gb:KarmaTrauttMishin:2012}%
  \BibitemOpen
  \bibfield{author}{%
  \bibinfo {author} {\bibfnamefont{A.}~\bibnamefont{Karma}}, \bibinfo {author}
  {\bibfnamefont{Z.~T.}\ \bibnamefont{Trautt}},\ and\ \bibinfo {author}
  {\bibfnamefont{Y.}~\bibnamefont{Mishin}},\ }%
  \bibfield{journal}{%
  \bibinfo {journal} {Phys. Rev. Lett.}\ }%
  \textbf{\bibinfo {volume} {109}},\ \bibinfo {pages} {095501} (\bibinfo {year}
  {2012})\BibitemShut{NoStop}%
\bibitem{gbseg:Seidman:2002}%
  \BibitemOpen
  \bibfield{author}{%
  \bibinfo {author} {\bibfnamefont{D.}~\bibnamefont{Seidman}},\ }%
  \bibfield{journal}{%
  \bibinfo {journal} {Annu. Rev. Mater. Res.}\ }%
  \textbf{\bibinfo {volume} {32}},\ \bibinfo {pages} {235} (\bibinfo {year}
  {2002})\BibitemShut{NoStop}%
\bibitem{imdrag:LuckeDetert:1957}%
  \BibitemOpen
  \bibfield{author}{%
  \bibinfo {author} {\bibfnamefont{K.}~\bibnamefont{L\"{u}cke}}\ and\ \bibinfo
  {author} {\bibfnamefont{K.}~\bibnamefont{Detert}},\ }%
  \bibfield{journal}{%
  \bibinfo {journal} {Acta Met.}\ }%
  \textbf{\bibinfo {volume} {5}},\ \bibinfo {pages} {628} (\bibinfo {year}
  {1957})\BibitemShut{NoStop}%
\bibitem{imdrag:Cahn:1962}%
  \BibitemOpen
  \bibfield{author}{%
  \bibinfo {author} {\bibfnamefont{J.~W.}\ \bibnamefont{Cahn}},\ }%
  \bibfield{journal}{%
  \bibinfo {journal} {Acta Met.}\ }%
  \textbf{\bibinfo {volume} {10}},\ \bibinfo {pages} {789} (\bibinfo {year}
  {1962})\BibitemShut{NoStop}%
\bibitem{gbm:LuckeStuwe:1971}%
  \BibitemOpen
  \bibfield{author}{%
  \bibinfo {author} {\bibfnamefont{K.}~\bibnamefont{L{$\ddot{\text{u}}$}cke}}\
  and\ \bibinfo {author} {\bibfnamefont{H.-P.}\
  \bibnamefont{St{$\ddot{\text{u}}$}we}}\ }%
  \textbf{\bibinfo {volume} {19}},\ \bibinfo {pages} {1087} (\bibinfo {year}
  {1971})\BibitemShut{NoStop}%
\bibitem{intpot:HepburnAckland:2008}%
  \BibitemOpen
  \bibfield{author}{%
  \bibinfo {author} {\bibfnamefont{D.~J.}\ \bibnamefont{Hepburn}}\ and\
  \bibinfo {author} {\bibfnamefont{G.}~\bibnamefont{Ackland}},\ }%
  \bibfield{journal}{%
  \bibinfo {journal} {Phys. Rev. B}\ }%
  \textbf{\bibinfo {volume} {78}},\ \bibinfo {pages} {165115} (\bibinfo {year}
  {2008})\BibitemShut{NoStop}%
\bibitem{atsim:FrenkelSmit:2002}%
  \BibitemOpen
  \bibfield{author}{%
  \bibinfo {author} {\bibfnamefont{D.}~\bibnamefont{Frenkel}}\ and\ \bibinfo
  {author} {\bibfnamefont{B.}~\bibnamefont{Smit}},\ }%
  \emph{\bibinfo {title} {Understanding molecular simulation: {F}rom algorithms
  to applications}},\ \bibinfo {edition} {2nd}\ ed.,\ Computational science\
  (\bibinfo {publisher} {Academic},\ \bibinfo {address} {San Diego},\ \bibinfo
  {year} {2002})\ p.\ \bibinfo {pages} {638}\BibitemShut{NoStop}%
\bibitem{gbm:TrauttUpmanyuKarma:2006}%
  \BibitemOpen
  \bibfield{author}{%
  \bibinfo {author} {\bibfnamefont{Z.~T.}\ \bibnamefont{Trautt}}, \bibinfo
  {author} {\bibfnamefont{M.}~\bibnamefont{Upmanyu}},\ and\ \bibinfo {author}
  {\bibfnamefont{A.}~\bibnamefont{Karma}},\ }%
  \bibfield{journal}{%
  \bibinfo {journal} {Science}\ }%
  \textbf{\bibinfo {volume} {27}},\ \bibinfo {pages} {632} (\bibinfo {year}
  {2006})\BibitemShut{NoStop}%
\bibitem{gbm:Schonfelder:1997}%
  \BibitemOpen
  \bibfield{author}{%
  \bibinfo {author}
  {\bibfnamefont{B.}~\bibnamefont{Sch{$\ddot{\text{o}}$}nfelder}}, \bibinfo
  {author} {\bibfnamefont{D.}~\bibnamefont{Wolf}}, \bibinfo {author}
  {\bibfnamefont{S.~R.}\ \bibnamefont{Phillpot}},\ and\ \bibinfo {author}
  {\bibfnamefont{M.}~\bibnamefont{Furtkamp}},\ }%
  \bibfield{journal}{%
  \bibinfo {journal} {Int. Sci.}\ }%
  \textbf{\bibinfo {volume} {5}},\ \bibinfo {pages} {245} (\bibinfo {year}
  {1997})\BibitemShut{NoStop}%
\bibitem{gbm:ZhangMendelev:2004}%
  \BibitemOpen
  \bibfield{author}{%
  \bibinfo {author} {\bibfnamefont{H.}~\bibnamefont{Zhang}}, \bibinfo {author}
  {\bibfnamefont{M.~I.}\ \bibnamefont{Mendelev}},\ and\ \bibinfo {author}
  {\bibfnamefont{D.~J.}\ \bibnamefont{Srolovitz}},\ }%
  \bibfield{journal}{%
  \bibinfo {journal} {Acta Mater.}\ }%
  \textbf{\bibinfo {volume} {52}},\ \bibinfo {pages} {2569} (\bibinfo {year}
  {2004})\BibitemShut{NoStop}%
\bibitem{gb:CahnLarche:1978}%
  \BibitemOpen
  \bibfield{author}{%
  \bibinfo {author} {\bibfnamefont{J.~W.}\ \bibnamefont{Cahn}}\ and\ \bibinfo
  {author} {\bibfnamefont{F.~C.}\ \bibnamefont{{Larch\'e}}},\ }%
  \bibfield{journal}{%
  \bibinfo {journal} {Acta Met.}\ }%
  \textbf{\bibinfo {volume} {26}},\ \bibinfo {pages} {1579} (\bibinfo {year}
  {1978})\BibitemShut{NoStop}%
\bibitem{imdrag:MendelevSrolovitz:2002}%
  \BibitemOpen
  \bibfield{author}{%
  \bibinfo {author} {\bibfnamefont{M.~I.}\ \bibnamefont{Mendelev}}\ and\
  \bibinfo {author} {\bibfnamefont{D.~J.}\ \bibnamefont{Srolovitz}},\ }%
  \bibfield{journal}{%
  \bibinfo {journal} {Mod. Sim. Mat. Sci. Eng.}\ }%
  \textbf{\bibinfo {volume} {10}},\ \bibinfo {pages} {R79} (\bibinfo {year}
  {2002})\BibitemShut{NoStop}%
\bibitem{imdrag:MendelevSrolovitzAckland:2005}%
  \BibitemOpen
  \bibfield{author}{%
  \bibinfo {author} {\bibfnamefont{M.~I.}\ \bibnamefont{Mendelev}}, \bibinfo
  {author} {\bibfnamefont{D.~J.}\ \bibnamefont{Srolovitz}}, \bibinfo {author}
  {\bibfnamefont{G.~J.}\ \bibnamefont{Ackland}},\ and\ \bibinfo {author}
  {\bibfnamefont{S.}~\bibnamefont{Han}},\ }%
  \bibfield{journal}{%
  \bibinfo {journal} {J. Mater. Res.}\ }%
  \textbf{\bibinfo {volume} {20}},\ \bibinfo {pages} {208} (\bibinfo {year}
  {2005})\BibitemShut{NoStop}%
\bibitem{imdrag:ZepedaRuizGilmer:2006}%
  \BibitemOpen
  \bibfield{author}{%
  \bibinfo {author} {\bibfnamefont{L.~A.}\ \bibnamefont{{Zepeda-Ruiz}}},
  \bibinfo {author} {\bibfnamefont{G.~H.}\ \bibnamefont{Gilmer}}, \bibinfo
  {author} {\bibfnamefont{B.}~\bibnamefont{Sadigh}}, \bibinfo {author}
  {\bibfnamefont{A.}~\bibnamefont{Caro}}, \bibinfo {author}
  {\bibfnamefont{T.}~\bibnamefont{Oppelstrup}},\ and\ \bibinfo {author}
  {\bibfnamefont{A.~V.}\ \bibnamefont{Hamza}},\ }%
  \bibfield{journal}{%
  \bibinfo {journal} {Appl. Phys. Lett.}\ }%
  \textbf{\bibinfo {volume} {87}},\ \bibinfo {pages} {231904} (\bibinfo {year}
  {2005})\BibitemShut{NoStop}%
\bibitem{gbm:Kappes:2008}%
  \BibitemOpen
  \bibfield{author}{%
  \bibinfo {author} {\bibfnamefont{B.~B.}\ \bibnamefont{Kappes}},\ }%
  \emph{\bibinfo {title} {The thermokinetic properties of high angle aluminum
  grain boundaries}},\ Ph.D. thesis,\ \bibinfo {school} {Colorado School of
  Mines} (\bibinfo {year} {2008})\BibitemShut{NoStop}%
\bibitem{Note1}%
  \BibitemOpen
  \bibinfo {note} {Elastic anisotropy can be important for the loaded boundary
  as the chemical potential of interstitial carbon is sensitive to stress
  distribution within the grain boundary region}\BibitemShut{NoStop}%
\bibitem{imdrag:RoyBauer:1975}%
  \BibitemOpen
  \bibfield{author}{%
  \bibinfo {author} {\bibfnamefont{A.}~\bibnamefont{Roy}}\ and\ \bibinfo
  {author} {\bibfnamefont{C.~L.}\ \bibnamefont{Bauer}},\ }%
  \bibfield{journal}{%
  \bibinfo {journal} {Acta Met.}\ }%
  \textbf{\bibinfo {volume} {23}},\ \bibinfo {pages} {957} (\bibinfo {year}
  {1975})\BibitemShut{NoStop}%
\bibitem{imdrag:MendelevSrolovitzWE:2001}%
  \BibitemOpen
  \bibfield{author}{%
  \bibinfo {author} {\bibfnamefont{M.~I.}\ \bibnamefont{Mendelev}}, \bibinfo
  {author} {\bibfnamefont{D.~J.}\ \bibnamefont{Srolovitz}},\ and\ \bibinfo
  {author} {\bibfnamefont{W.}~\bibnamefont{E}},\ }%
  \bibfield{journal}{%
  \bibinfo {journal} {Phil. Mag. A}\ }%
  \textbf{\bibinfo {volume} {81}},\ \bibinfo {pages} {2243} (\bibinfo {year}
  {2001})\BibitemShut{NoStop}%
\bibitem{imdrag:KorzhenevskiiBauschSchmitz:2006}%
  \BibitemOpen
  \bibfield{author}{%
  \bibinfo {author} {\bibfnamefont{A.~L.}\ \bibnamefont{Korzhenevskii}},
  \bibinfo {author} {\bibfnamefont{R.}~\bibnamefont{Bausch}},\ and\ \bibinfo
  {author} {\bibfnamefont{R.}~\bibnamefont{Schmitz}},\ }%
  \bibfield{journal}{%
  \bibinfo {journal} {Acta Mater.}\ }%
  \textbf{\bibinfo {volume} {54}},\ \bibinfo {pages} {1595} (\bibinfo {year}
  {2006})\BibitemShut{NoStop}%
\bibitem{digm:CahnPanBalluffi:1979}%
  \BibitemOpen
  \bibfield{author}{%
  \bibinfo {author} {\bibfnamefont{J.~W.}\ \bibnamefont{Cahn}}, \bibinfo
  {author} {\bibfnamefont{J.~D.}\ \bibnamefont{Pan}},\ and\ \bibinfo {author}
  {\bibfnamefont{R.~W.}\ \bibnamefont{Balluffi}},\ }%
  \bibfield{journal}{%
  \bibinfo {journal} {Scripta Metall.}\ }%
  \textbf{\bibinfo {volume} {13}},\ \bibinfo {pages} {503} (\bibinfo {year}
  {1979})\BibitemShut{NoStop}%
\bibitem{digm:HillertPurdy:1978}%
  \BibitemOpen
  \bibfield{author}{%
  \bibinfo {author} {\bibfnamefont{M.}~\bibnamefont{Hillert}}\ and\ \bibinfo
  {author} {\bibfnamefont{G.~R.}\ \bibnamefont{Purdy}},\ }%
  \bibfield{journal}{%
  \bibinfo {journal} {Acta Met.}\ }%
  \textbf{\bibinfo {volume} {26}},\ \bibinfo {pages} {333} (\bibinfo {year}
  {1978})\BibitemShut{NoStop}%
\bibitem{digm:Handwerker:1988}%
  \BibitemOpen
  \bibfield{author}{%
  \bibinfo {author} {\bibfnamefont{C.}~\bibnamefont{Handwerker}},\ }%
  in\ \emph{\bibinfo {booktitle} {Diffusion phenomena in thin films and
  microelectronic materials}},\ Vol.~\bibinfo {volume} {26},\ \bibinfo {editor}
  {edited by\ \bibinfo {editor} {\bibfnamefont{D.}~\bibnamefont{Gupta}}\ and\
  \bibinfo {editor} {\bibfnamefont{P.~S.}\ \bibnamefont{Ho}}}\ (\bibinfo
  {publisher} {Noyes Publications},\ \bibinfo {address} {Park Ridge, NJ},\
  \bibinfo {year} {1988})\ pp.\ \bibinfo {pages}
  {245--322}\BibitemShut{NoStop}%
\bibitem{digm:Penrose:2004}%
  \BibitemOpen
  \bibfield{author}{%
  \bibinfo {author} {\bibfnamefont{O.}~\bibnamefont{Penrose}},\ }%
  \bibfield{journal}{%
  \bibinfo {journal} {Acta Mater.}\ }%
  \textbf{\bibinfo {volume} {52}},\ \bibinfo {pages} {3901–3910} (\bibinfo
  {year} {2011})\BibitemShut{NoStop}%
  \BibitemOpen
  \BibitemShut{NoStop}%
\end{thebibliography}
\end{document}